\documentclass[sigconf]{acmart}

\renewcommand\footnotetextcopyrightpermission[1]{} 

\usepackage{amsmath,amsfonts,bm}

\def\eqref#1{equation~\ref{#1}}

\def\1{\bm{1}}

\def\vt{{\bm{t}}}

\def\vx{{\bm{x}}}

\DeclareMathAlphabet{\mathsfit}{\encodingdefault}{\sfdefault}{m}{sl}
\SetMathAlphabet{\mathsfit}{bold}{\encodingdefault}{\sfdefault}{bx}{n}
\newcommand{\tens}[1]{\bm{\mathsfit{#1}}}

\def\tB{{\tens{B}}}

\usepackage{pifont}
\usepackage{amsmath}
\usepackage{amsfonts}
\usepackage{graphicx}
\usepackage{textcomp}
\usepackage{xcolor}
\usepackage{multirow}
\usepackage{soul}
\usepackage{tikz}
\usepackage{booktabs}
\usepackage{algorithm}
\usepackage{algpseudocode}
\usepackage{colortbl}
\usepackage{bm}
\usepackage{float}
\usepackage{filecontents}
\usepackage{CJK}
\usepackage{subcaption}
\usepackage{algorithm}
\usepackage{algorithmicx}

\usepackage{pifont}

\begin{document}

\title{Threshold Breaker: Can Counter-Based RowHammer Prevention Mechanisms Truly Safeguard DRAM?}

 \author{Ranyang Zhou$^\dagger$, Jacqueline Liu$^\ddagger$, Sabbir Ahmed$^\ddagger$, Nakul Kochar$^\dagger$,\\ Adnan Siraj Rakin$^\ddagger$, Shaahin Angizi$^\dagger$}
\affiliation{\vspace{0.5em}
	\institution{$^\dagger$ Department of Electrical and Computer Engineering, New Jersey Institute of Technology, Newark, NJ, USA}
   \country{}
	\institution{$^\ddagger$ Department of Computer Science, State University of New York at Binghamton, NY, USA}{}
}
\email{arakin@binghamton.edu, 
 shaahin.angizi@njit.edu}

\begin{abstract}
This paper challenges the existing victim-focused counter-based RowHammer detection mechanisms by experimentally demonstrating a novel multi-sided fault injection attack technique called Threshold Breaker. This mechanism can effectively bypass the most advanced counter-based defense mechanisms by soft-attacking the rows at a farther physical distance from the target rows. While no prior work has demonstrated the effect of such an attack, our work closes this gap by systematically testing 128 real commercial DDR4 DRAM products and reveals that the Threshold Breaker affects various chips from major DRAM manufacturers. As a case study, we compare the performance efficiency between our mechanism and a well-known double-sided attack by performing adversarial weight attacks on a modern Deep Neural Network (DNN). The results demonstrate that the Threshold Breaker can deliberately deplete the intelligence of the targeted DNN system while DRAM is fully protected. 
\end{abstract} \vspace{-1em}





\maketitle
\pagestyle{plain} 

\section{Introduction}
To guarantee system reliability and security, it is imperative to uphold memory isolation, ensuring that accessing a memory address does not lead to unintended side effects on data stored in other addresses. 
However, with the aggressive technology scaling, accessing (reading) a  Dynamic Random Access Memory (DRAM) cell disrupts the stored charge of other physically adjacent DRAM cells, causing bit-flips. Bit-flips have been enabled mainly due to a manifestation of a DRAM cell-to-cell interference and failure mechanism called RowHammer (RH) \cite{kim2014flipping,mutlu2023fundamentally,mutlu2019rowhammer,hassan2017softmc} as depicted in Fig. \ref{DRAM}. RH attack is conducted when a malicious process activates and pre-charges a specific row (i.e., aggressor row) repeatedly to a certain threshold ($T_{RH}$) to induce bit-flips on immediate nearby rows (i.e., victim rows). 
Unfortunately, by scaling down the size of DRAM chips in the modern manufacturing process, DRAM becomes increasingly more vulnerable to RH bit-flip. For example, the attacker needs $\sim$4.5$\times$ fewer hammer counts on LPDDR4 as opposed to DDR3 \cite{woo2022scalable}.
At the application level, recent studies show that an adversary can identify and manipulate a small number of vulnerable bits of off-the-shelf well-trained Deep Neural Network (DNN) weight parameters to significantly compromise the output accuracy \cite{hong2019terminal,rakin2019bit,angizi2018cmp}.

Addressing RH errors necessitates the implementation of more robust Error Correction Code (ECC) techniques, which come at the cost of excessive energy consumption, reduced performance, higher cost, and capacity overhead \cite{mutlu2019rowhammer,lee2019twice,angizi2020pim}.
The standard RH mitigation approach used by system manufacturers such as Apple \cite{Apple} and HP \cite{HP} is to increase the refresh rate which imposes a humongous power consumption and can be easily compromised \cite{mutlu2019rowhammer,rakin2021deep}. Intel's pTRR \cite{kaczmarski2014thoughts} and several research works propose to proactively count the number of row activations, i.e., Hammer Counts (HC) by maintaining an array of counters in either the memory controller \cite{bains2015method} or in the DRAM chips themselves \cite{kim2014architectural,qureshi2022hydra,seyedzadeh2016counter,frigo2020trrespass,wang2019detect}. Memory controller keeps the HC track and refreshes victim rows when the number of row activations issued to the DRAM exceeds Maximum Activate Count (MAC) threshold ($T_{MAC}$) which is typically saved on the Serial Presence Detect (SPD) chip within the DRAM module \cite{frigo2020trrespass}. Time Window Counter (TWiCe) \cite{lee2019twice} limits the counter entries per DRAM bank and imposes no performance overhead and less than 0.7\% energy and area overheads. 

In this paper, \textit{we counterargue the existing counter-based RH prevention mechanisms will be effective in protecting DRAM}. The key question this paper will answer is that \textit{Can we conduct an effective targeted bit-flip attack, with smaller HC than a double-sided RH attack, on the physically adjacent rows to bypass the MAC?} The main contributions of this paper are as follows: 

(1) For the first time, we develop a new multi-sided RH fault injection technique called Threshold Breaker that can manipulate DRAM data by soft-attacking the rows at a farther physical distance from the target rows, bypassing the MAC;  

(2) We experimentally analyze and verify the impact of the Threshold Breaker on 128 DDR4 DRAM chips across various DRAM manufacturers, namely Samsung, Micron, etc., with counter-based RH protection mechanisms enabled;


(3) As a case study, we demonstrate that leveraging an adversarial weight attack algorithm~\cite{rakin2019bit}, Threshold Breaker can compromise the performance of a modern DNN model bypassing MAC, whereas the popular double-sided RH attack fails. \vspace{-0.4em}

\section{Overview}
\begin{figure}[t]
\begin{center}\vspace{-0.5em}
\begin{tabular}{c}
\includegraphics [width=0.97\linewidth]{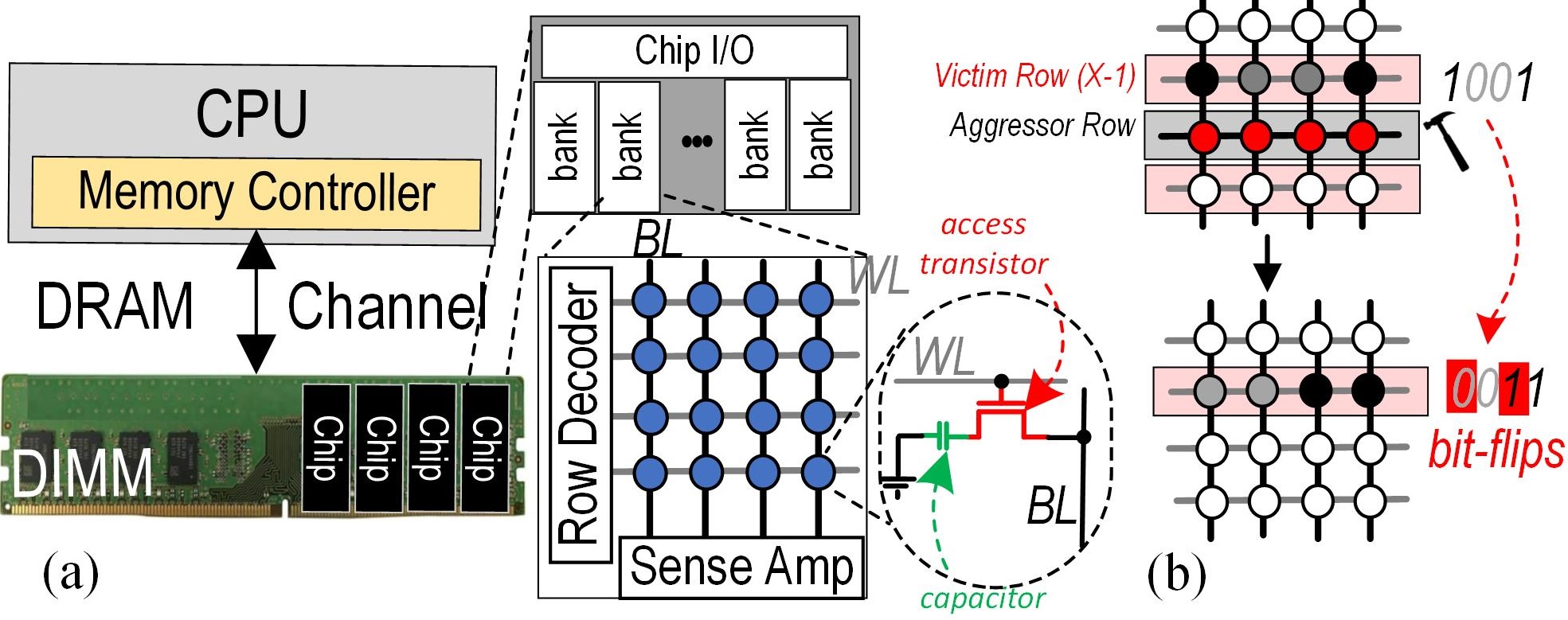}\vspace{-0.4em}
 \end{tabular} \vspace{-0.8em}
\caption{(a) Hierarchical organization of DRAM, (b) RowHammer attack \cite{kim2014flipping}.}\vspace{-1em}
\label{DRAM} \vspace{-0.2em}
\end{center}
\end{figure}

\textbf{DRAM Organization \& Commands.} The DRAM chip is a hierarchical structure consisting of several memory banks, as shown in Fig. \ref{DRAM}(a). Each bank comprises 2D sub-arrays of memory bit-cells virtually ordered in memory matrices, with billions of DRAM cells on modern chips. Each DRAM bit-cell consists of a capacitor and an access transistor. The charge status of the bit-cell's capacitor is used to represent binary ``1'' or ``0'' \cite{seshadri2017ambit,zhou2022red,angizi2019graphide,angizi2019redram}. In idle mode, the memory controller turns off all enabled DRAM rows by sending the Precharge (\texttt{PRE}) command on the command bus. This will precharge the Bit-Line (BL) voltage to $\frac{V_{DD}}{2}$. In the active mode,
 the memory controller will send an Activate (\texttt{ACT)} command to
the DRAM 
module to activate the Word-Line (WL). Then, all DRAM cells connected to the WL share their charges with the corresponding BL. Through this process, BL voltage deviates from the precharged $\frac{V_{DD}}{2}$. The sense amplifier then senses this deviation and amplifies it to $V_{DD}$ or 0 in the row buffer. The memory controller can then send read (\texttt{RD})/write (\texttt{WR}) commands to transfer data from/to the sense amplifier array \cite{zhang2021max,zhang2023aligner,angizi2019accelerating}.

\noindent\textbf{DRAM Timing Parameters.}
In the context of DRAM standards, a comprehensive array of timing parameters is established, with each parameter prescribing the minimum temporal separation between two successive DRAM commands to uphold seamless operational integrity. The most basic parameter is the clock cycle ($t_{CK}$) used to measure all parameters. Row Active Time (\textbf{$t_{RAS}$}) encompasses the temporal window demarcating an \texttt{ACT} command and the subsequent \texttt{PRE} command. During this prescribed $t_{RAS}$ interval, the restoration of charge within the DRAM cells on the open DRAM row is effectuated to ensure optimal performance. Row Precharge Time (\textbf{$t_{RP}$}) signifies the temporal gap between the issuance of a \texttt{PRE} command and the subsequent \texttt{ACT} command. The imposition of $t_{RP}$ is instrumental in closing the open WL and initiating the pre-charging of the DRAM BLs to the voltage level of $\frac{V_{DD}}{2}$. Retention time in DRAM refers to the duration for which a memory cell can hold its stored data without requiring a refresh operation. It can be influenced by various factors, including the density of cells, electromagnetic interference, and so on. The critical timing parameters are fundamental in ensuring the reliable and efficient operation of DRAM modules across different standards. In the RH model, the retention time of certain victim rows may experience a substantial reduction. The Refresh Window $t_{REFW}$ is essentially the interval within which all DRAM cells must be refreshed to prevent data loss or corruption.

\noindent\textbf{RH in DDR4 \& Protection Mechanisms.}
Kim et al. \cite{kim2014flipping} were the pioneers in conducting an extensive study on the characteristics of RH bit flips in DDR3 modules. They observed that approximately 85\% of the tested modules were susceptible to RH attack. Therefore, the majority of earlier RH research is centered on DDR3 systems \cite{seaborn2015exploiting}. With the prospect of having an RH-less landscape, DDR4 modules have been introduced. While there are documented instances of RH on DDR4 chips in previous studies \cite{lipp2020nethammer,gruss2018another}, these findings pertain to earlier generations of DDR4.  To the best of our knowledge, the only recent and established work exploring the multi-sided fault injection model is TRRespass \cite{frigo2020trrespass}. 

Multiple software and hardware mitigation mechanisms have been proposed to reduce the impact of RH-based attacks \cite{kim2014flipping,zhou2022lt,marazzi2022protrr,zhou2023dnn}. 
The hardware-based research efforts can be classified into two categories, i.e., \textit{victim-focused} mechanism with probabilistic refreshing (e.g., PRA \cite{kim2014architectural}, PARA \cite{kim2014flipping}, ProHIT \cite{son2017making}, ProTRR \cite{marazzi2022protrr}) and \textit{aggressor-focused} mechanism by counting activations (e.g.,  TRR \cite{hassan2021uncovering}, Hydra \cite{qureshi2022hydra}, CBT \cite{seyedzadeh2018mitigating}, Panopticon \cite{bennett2021panopticon}, CRA \cite{kim2014architectural}, TWiCe \cite{lee2019twice}, Graphene \cite{park2020graphene}, Mithril \cite{kim2022mithril}).
The system manufacturers tend to follow the mechanisms that explicitly detect RH conditions and intervene, such as increasing refresh rates and access counter-based approaches. 
Along this line, Target Row Refresh (TRR) \cite{frigo2020trrespass} and counter-based detection methods \cite{kim2014architectural,qureshi2022hydra,seyedzadeh2016counter} require add-on hardware to calculate rows' activation and record it to other fast-read-memory (SRAM \cite{lee2019twice}/CAM \cite{park2020graphene}). The controller will then refresh the target row if the number reaches MAC \cite{frigo2020trrespass}.
TWiCe \cite{lee2019twice} is a per-row counter-based RH prevention solution based on the idea that the number of ACTs within $t_{REFW}$ is limited. Instead of detecting the rows, TWiCe only checks the number of \texttt{ACT}s. However, inserting a counter for each memory row imposes a substantial burden both from latency and power consumption perspectives \cite{seyedzadeh2016counter}. To tackle this issue, recent works \cite{seyedzadeh2016counter} consider the storage of counters in a dedicated section of DRAM or use a set-associative counter cache implemented within the memory controller to enhance the efficiency of accessing frequently utilized counters \cite{kim2014architectural}.
CAT \cite{seyedzadeh2018mitigating} is a counter-based solution that keeps track of the number of \texttt{ACT}s performed on a set of DRAM rows and initiates a refresh operation for the entire group of rows, once the HC reaches the MAC. The counter-based solutions have been enabled by adding a new DRAM command called Nearby Row Refresh (NRR) \cite{lee2019twice,park2020graphene}
that will be issued to refresh the relevant victim rows.
The JEDEC standard outlines three potential configurations for the MAC value: (1) unlimited, if the DRAM module claims to be RH-free; (2) untested, if the DRAM module has not undergone post-production inspection; or (3) $T_{MAC}$ indicating the specific number of \texttt{ACT}s the DRAM module can withstand (e.g., 1M). It has been revealed \cite{frigo2020trrespass} that, irrespective of the DRAM manufacturer, the majority of DDR4 modules assert unlimited MAC value.
\vspace{-1em}

\section{Threshold Breaker}
To enhance the defensive capabilities of DRAM modules, it is necessary to adopt an attacker's perspective, enabling a deeper comprehension of potential threats and more effective countermeasures. Existing counter-based RH prevention frameworks come with distinct challenges, specifically pertaining to their scope and thresholds. From an attacker's perspective, we can articulate three essential directions to defeating counter-based frameworks: $(i)$ Broaden the attack area as extensively as possible to make the detection more complicated; $(ii)$ Leverage various attack patterns such as side-kick aggressors or many-sided attacks \cite{frigo2020trrespass,saroiu2022price}; and $(iii)$ Reduce the HC if possible to fool the system by not being detected. Our objective in this section is to explore the third direction to elevate the cost of counter-based defense in DDR4 modules and, in the end, overcome established mitigation techniques.

\begin{figure}[t]
\begin{center}
\begin{tabular}{c}
\includegraphics [width=1\linewidth]{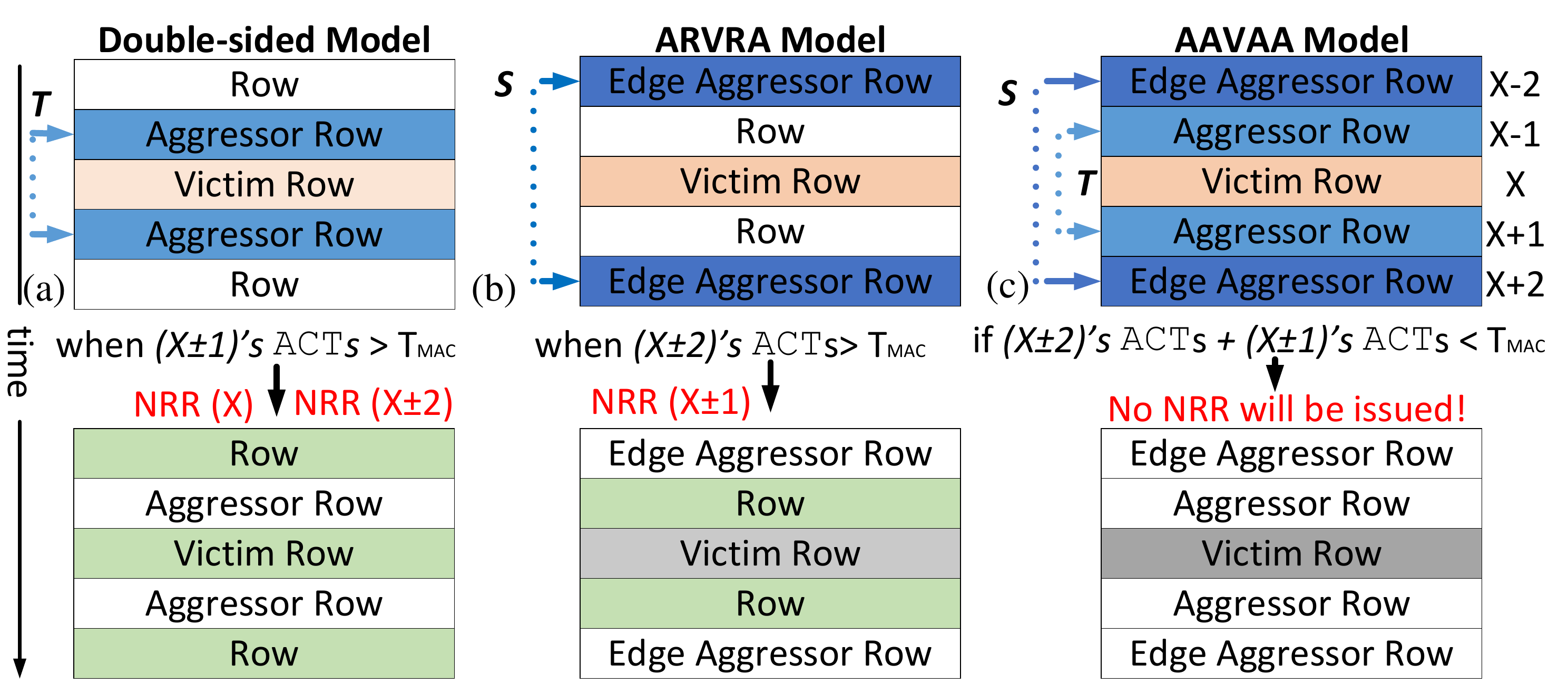}\vspace{-0.4em}
 \end{tabular} \vspace{-0.9em}
\caption{RowHammer attack models (top) and refreshed rows after NRR commands (bottom): (a) Double-sided model, (b) under-test ARVRA model, (c) Proposed AAVAA model.}
\label{RHmodel}
\vspace{-1.5em}
\end{center}
\end{figure}

\begin{figure}[h]
\begin{center}\vspace{-1.4em}
\begin{tabular}{c}
\includegraphics [width=0.99\linewidth]{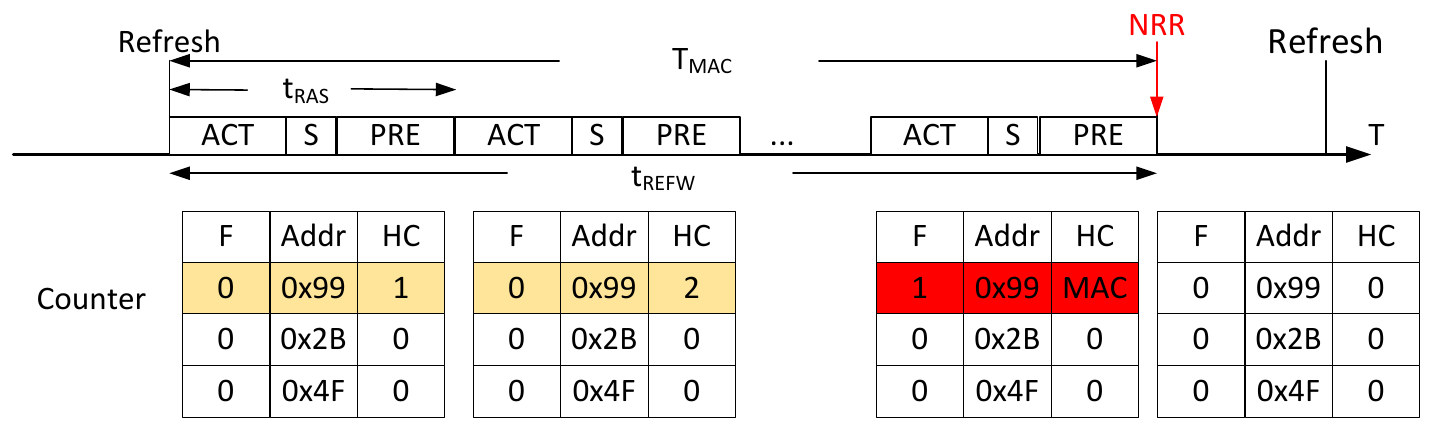}\vspace{-0.4em}
 \end{tabular} \vspace{-1.6em}
\caption{Timing of RH attack.}
\label{timing}
\vspace{-1.3em}
\end{center}
\end{figure}

Traditional fault injection models such as double-sided and sandwich attacks can be effectively defended \cite{saroiu2022price,kaczmarski2014thoughts} by counter-based frameworks. As shown in Fig. \ref{RHmodel}(a)-top, the double-sided RH model mainly affects the victim rows with two aggressors X$\pm$1. While there are three victim rows in this model, the primary focus of this approach is on victim row X, as both aggressor rows simultaneously exert a significant influence on it. Subsequent testing allows us to establish a range of aggressor rows' HCs, denoted by $T$, that effectively quantifies the vulnerability levels of the victim rows. The lower and higher boundaries of $T$ correspond to the respective thresholds where the victim row first exhibits bit-flips and where the victim row is entirely reversed due to the attacks, respectively. Hence, defense mechanisms will easily identify anomalous rows that have been activated significantly more frequently than typical rows. As discussed, such defenses establish distinct thresholds depending on the manufacturer of the chips. If the defense mechanism properly detects that the row X$\pm$1 reaches the $T_{MAC}$, the NRR will refresh row X and X$\pm$2 as shown in Fig. \ref{RHmodel}(a)-bottom. Fig. \ref{timing} shows the timing for such an RH attack. Assuming RH is implemented on the row 0x99. F is a flag used to decide whether to issue an NRR command or not. When HC for the row surpasses MAC, which means $t_{RAS}$$\times$ HC $\geq$$T_{MAC}$, the memory controller considers an NRR operation for that row.
Common $t_{RAS}$ values for DDR4 memory modules could range from around 36 to 48 $t_{CK}$ \cite{choi2020reducing}, but these values can differ based on the module's speed rating (e.g., DDR4-2133, DDR4-2400, DDR4-3200, etc.). The duration of a clock cycle for DDR4-2400 memory can be calculated as $t_{RAS}=\frac{1}{2400MT/s}$. In our design, every $t_{RAS}$ consists of three parts: \texttt{ACT}, \texttt{Sleep(S)}, and \texttt{PRE}, where \texttt{Sleep(S)} is set to 5{$t_{CK}$}. 

\noindent\textbf{Multi-sided Fault Injection Model.}
The multiple-row fault injection model of Threshold Breaker herein represents a new concept rooted in the traditional double-sided attack model with an expanded set of attack vectors. Through the examination, our objective is to identify potential HCs to enable a so-called soft-attack, i.e., weaker attack with fewer aggressor rows' \texttt{ACT}, as a means to circumvent counter-based defenses regardless of DRAM controller implementation details in various DRAM chips. 

The under-test \textit{\ul{ARVRA RH model}} depicted in Fig. \ref{RHmodel}(b)-top is a straightforward variant of the double-sided model, in which the two edge aggressor rows (X$\pm$2) that are one row apart from the targeted victim row, hammer it $S$ times. Our working hypothesis is when (X$\pm$2)'s \texttt{ACT}s is greater than $T_{MAC}$,
ARVRA model can victimize all three sandwiched rows and may effectively flip the bits in the targeted victim row X. By issuing the NRR command, X$\pm$1 rows will be refreshed where the victim row remains flipped as shown in Fig. \ref{RHmodel}(b)-bottom.
The proposed \textit{\ul{AAVAA RH model}} aims to combine the double-sided model and ARVRA model in order to find a way to bypass the $T_{MAC}$. In this model, as shown in Fig. \ref{RHmodel}(c)-top, so-called edge aggressor rows (X$\pm$2) and typical aggressor rows (X$\pm$1) are soft-attacked/hammered $S$ and $T$ times, respectively. \textit{Our working hypothesis is that current DRAM modules exposed to AAVAA might be vulnerable to certain reduced hammering patterns by which (X$\pm$2)'s \texttt{ACT}s + (X$\pm$1)'s \texttt{ACT}s is less than $T_{MAC}$.} In this case, no counter-based technique will be able to figure out which row is victimized, and therefore no NRR command will be issued. TRRespass \cite{frigo2020trrespass} as the only established research on multiple-row RH fault injection presents a fuzzy approach to create diverse RH patterns through a completely black-box methodology. It demonstrates such patterns can effectively induce bit flips in DDR4 DRAM chips from all three major DRAM vendors. Contrary to TRRespass, the focus of Threshold Breaker is not on the pattern but on finding HC values specific to DDR4 modules to enable soft-attacking and unlock its attack potential.

\noindent\textbf{Algorithm.}
We develop the following \textit{CounterBypass} procedure to characterize various DDR4 DRAM modules and implement our proposed multi-sided fault injection model by enabling the control over the $S$ and $T$ parameters for different attack scenarios as shown in Algorithm \ref{alg}. We initialize the $row\_address$, $bank\_address$, and $column\_address$ at which the attack will be conducted in line-3. 
In the absence of data in the DRAM rows, the experiment becomes impractical. Therefore, we set all bits in aggressor rows to ``1'' and those in victim rows to ``0''. This initialization establishes a straightforward state, facilitating a clearer observation of the attack intensity. Given the disparity in bits within the same column of the aggressor and victim rows, a bit-flip could be induced. Subsequently, two RH operations are implemented to activate aggressor rows X$\pm$1 in a loop (lines 8-10). Same as this, another set of two aggressor rows X$\pm$2 are activated in a loop (lines 15-17).
Following the execution of all RH operations in lines 7-18, we retrieve the data from all rows. Ultimately, we ascertain the number of bit-flips by comparing the read data with the initially stored data. In this algorithm, designating $S$ or $T$ as zero signifies the transition to either the double-sided model or the ARVRA model; otherwise, it follows the AAVAA model. \vspace{-0.2em}

\begin{algorithm}[h]
        \caption{\small Multi-sided Fault Injection}
          \scalebox{0.65}{
    \begin{minipage}{3\linewidth}
        \begin{algorithmic}[1]
            \State $\textbf{Procedure: \textit{CounterBypass}}$
                \State $\textbf{Input} \hspace{4pt} S,T$ 
                \State $Allocate \hspace{4pt} row\_address, bank\_address, column\_address$ 
                \State $Load \hspace{4pt} Data\_pattern \hspace{4pt} $\&$ \hspace{4pt} Data\_pattern\_inv $ 
                \State $Aggressor\_row[row] \gets Data\_pattern\hspace{16pt}//We\hspace{4pt}assign\hspace{4pt}0xFFFFFFFF\hspace{4pt}to\hspace{4pt}aggressor\hspace{4pt}rows$
                \State $Victim\_row[row] \gets Data\_pattern\_inv\hspace{16pt}//We\hspace{4pt}assign\hspace{4pt}0x00000000\hspace{4pt}to\hspace{4pt}victim\hspace{4pt}rows$ 
                \State $\textbf{For}\hspace{4pt}(RowHammer\_cnt < S)\hspace{4pt}\textbf{do}$
                    \State $\hspace{16pt}\textbf{For}\hspace{4pt}(row\hspace{4pt} in\hspace{4pt}Row\_address)\hspace{4pt}\textbf{do}$
                    \State $\hspace{32pt}ACT \hspace{4pt} Aggressor\_row[row];\hspace{16pt}//Keep\hspace{4pt}hammering\hspace{4pt}Row\hspace{4pt}X\pm1$ 
                    \State $\hspace{32pt}PRE \hspace{4pt} Aggressor\_row[row];$ 
                    \State $\hspace{32pt}row \gets Row\_address+1;$ 
                \State $\hspace{16pt}RowHammer\_cnt+1;$
                \State $\textbf{For}\hspace{4pt}(RowHammer\_cnt < T)\hspace{4pt}\textbf{do}$
                    \State $\hspace{16pt}\textbf{For}\hspace{4pt}(row\hspace{4pt} in\hspace{4pt}Edge\_Aggressor\_row)\hspace{4pt}\textbf{do}$
                    \State $\hspace{32pt}ACT \hspace{4pt} Edge\_Aggressor\_row[row];\hspace{16pt}//Keep\hspace{4pt}hammering\hspace{4pt}Row\hspace{4pt}X\pm2$ 
                    \State $\hspace{32pt}PRE \hspace{4pt} Edge\_Aggressor\_row[row];$ 
                    \State $\hspace{32pt}row \gets Row\_address+1;$ 
                \State $\hspace{16pt}Edge\_RowHammer\_cnt+1;$
        \State $\textbf{For}\hspace{4pt}(row\hspace{4pt} in\hspace{4pt}Row\_address)\hspace{4pt}\textbf{do}$
                    \State $\hspace{16pt}READ \hspace{4pt} Aggressor\_row[row];$ 
                    \State $\hspace{16pt}PRE \hspace{4pt} Aggressor\_row[row];$ 
                    \State $\hspace{16pt}row \gets Row\_address+1;$ 
        \State $Receive\_Data(Platform);\hspace{16pt}//Write\hspace{4pt}data\hspace{4pt}back\hspace{4pt}to\hspace{4pt}hostPC$
        \State $Detect\_BitFlips(Victim\_Row)$
    \State \textbf{end} $\textbf{Procedure}$
        \end{algorithmic} 
         \end{minipage} \vspace{-3em}}
         \label{alg}
    \end{algorithm}

\vspace{-0.5em}
\section{Experimental Results}
\begin{figure}[t]
\begin{center}
\begin{tabular}{c}
\includegraphics [width=0.8\linewidth]{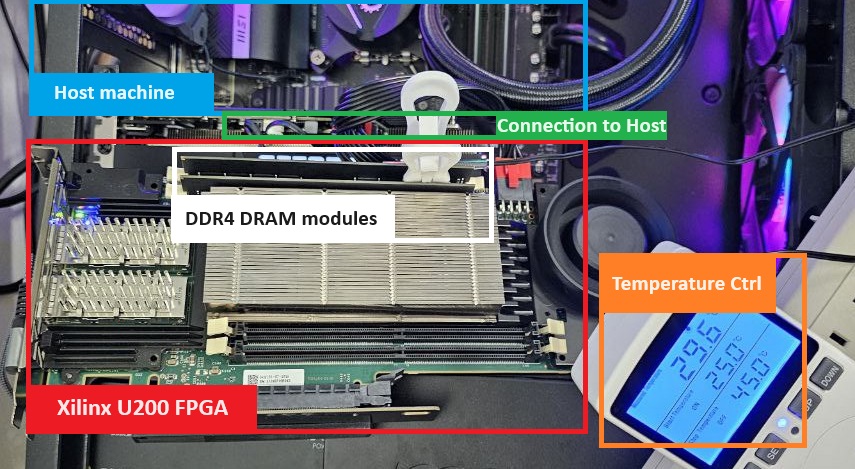}\vspace{-0.7em}
 \end{tabular} \vspace{-0.4em}
\caption{Our testing infrastructure for DDR4 modules.}
\label{frame}
\vspace{-2em}
\end{center}
\end{figure}

\textbf{Framework Setup \& Testing Infrastructure.} We test the DRAM chips by extensively modifying the DRAM-Bender \cite{olgun2023dram} to have a versatile FPGA-based DRAM attack exploration framework for DDR4 with an in-DRAM compiler API installed on our host machine. Our testing infrastructure, as shown in Fig. \ref{frame}, consists of the Alveo U200 Data Center Accelerator Card \cite{Alevo} as the FPGA that accepts DDR4 modules and runs the test programs based on Algorithm \ref{alg} by sending DDR4 command traces generated by the host machine. Besides, to have a fair comparison among various under-test DRAM chips, the temperature is kept below 30$^{\circ}$C with INKBIRDPLUS 1800W temperature controller.

\noindent\textbf{Minimizing Interference.} Before implementing the proposed attack scenario, DRAM refresh \cite{JEDEC} and rank-level ECC are disabled to minimize their interference with RH bit-flips. However, proprietary RH protection techniques (e.g., Target Row Refresh \cite{frigo2020trrespass,hassan2021uncovering}) are in place.

\noindent\textbf{Chips Tested.} 
To characterize and show the impact of Threshold Breaker, the experiments are conducted on a range of 128 commercialized DRAM chips from eight different manufacturers (mf.) as listed in Table \ref{DRAMchips} with various die densities and die revisions. 

\noindent\textbf{DNN Evaluation.} We took the data collected from three representative DRAM chips and evaluated a ResNet-34~\cite{he2016deep} model against the popular BFA attack algorithm~\cite{rakin2019bit} on the ImageNet~\cite{deng2009imagenet} dataset.
\vspace{-1em}

\begin{table}[h]
  \centering
  \caption{Under-test DRAM chips.} \vspace{-1.2em}
\scalebox{0.72}{
\begin{tabular}{lcccll}
\hline
\multicolumn{1}{c}{\textbf{Vendor}} & \textbf{\#Chips} & \textbf{Freq (MHz)}  & \textbf{Die rev.}    & \textbf{Org.} & \textbf{Date} \\ \hline
mf-A (Micron 16GB)                  & 16               & 2133                 &  B                    &x4               &  2126             \\
mf-B (ATECH 16GB)                 & 16               & 2933                 &A                      &x8               &  2597            \\ 
mf-C (Crucial 16 GB)                 & 16               & 3200                 &   C                   &x8              & N/A              \\ 
mf-D (Kingston 16GB)               & 16               & 2666    & G              &x8 & 2152              \\
mf-E (NEMIX 16GB)                 & 16               & 2133                 &B                      &x4               &   1733            \\ 
mf-F (SK Hynix 16GB)                & 16               & 2400                 &A                      &x8               & 1817              \\
mf-G (Patriot Viper 16GB)                 & 16               & 3600                 & C                    &x8               &  N/A             \\ 
mf-H (Samsung 16GB)                 & 16               & 2400                 & B                     & x8                 &  2053          \\

\hline
\end{tabular}}
  \label{DRAMchips} \vspace{-2em}
\end{table}

\subsection{Characterization Method and Observations}
\begin{figure*}[t]
\begin{center}
\begin{tabular}{c}
\includegraphics [width=0.9\linewidth]{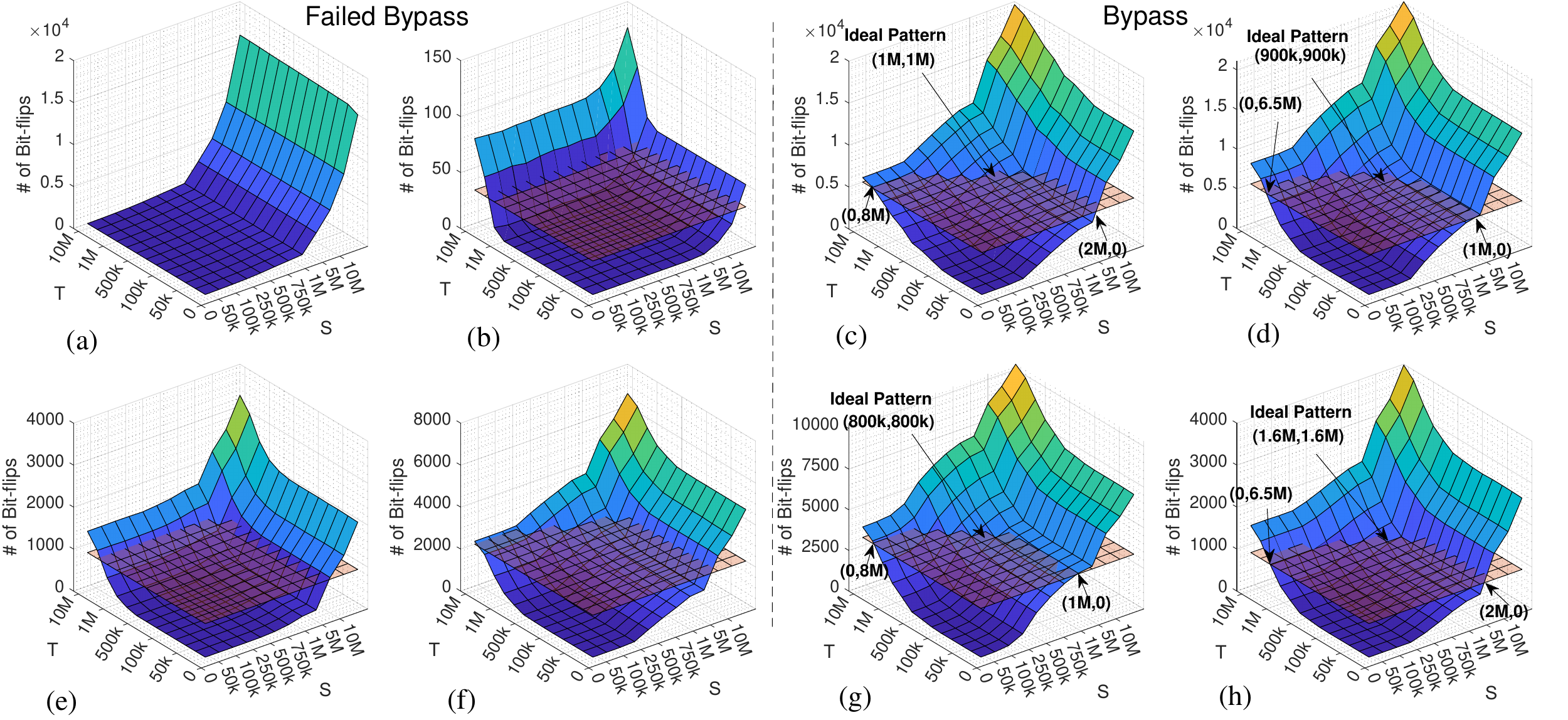}\vspace{-0.4em}
 \end{tabular} \vspace{-1.4em}
\caption{Experimental results demonstrating the number of bit-flips with (S,T) configuration on (a) mf-A, (b) mf-B, (c) mf-C, (d) mf-D, (e) mf-E, (f) mf-F, (g) mf-G, (h) mf-H.}
\label{results1}
 \vspace{-0.5em}
\end{center}
\end{figure*}

To study the effectiveness of Threshold Breaker on read disturbance, we first comprehensively analyze the AAVAA attack model on various ($S$, $T$) configuration sets. Please note that ARVRA is a sub-set of AAVAA whereby $T$ = 0. The characterization method remains consistent for DRAM modules from eight distinct manufacturers. It includes incrementing both $S$ and $T$ HCs to assess the effects of all conceivable combinations. The 3-D surface plots presented in Fig. \ref{results1} reveal distinct characteristics for each design, which an attacker can exploit to effectively bypass any counter-based defense mechanism. 
To facilitate understanding of results, a 2-D plane shows a fixed number of bit-flips incorporated within Fig. \ref{results1}(b)-(h).

We systematically examine 16 chips sourced from every manufacturer through a rigorous testing process. Our findings reveal that the performance differentials among these chips are minimal, with variations consistently below the 5\% threshold, attesting to their uniform quality and reliability. This implies that the test outcomes for individual chips are indicative of the overall performance of this manufacturer's chips. In this scenario, as we acquire multiple samples of identical chips, we compute the average test data to create a representative plot, ensuring a more accurate depiction of the manufacturer's chip characteristics.

\fbox{\begin{minipage}{25em}
\textbf{Obs.\#1.} 
The individual chip evaluation performance from the same manufacturer is fairly consistent with a minimal level of variation for independent rounds of attacks.
\end{minipage}}

The simultaneous testing of three models is achievable by manipulating the patterns of $S$ and $T$, as mentioned earlier. Initially, the data from the double-sided model and ARVRA model aligns with plots on the X-Z and Y-Z coordinates. It becomes evident that as the value of $S$ increases along the X-Z coordinates, the rate of bit-flips surpasses that observed along the Y-Z coordinates. Additionally, when $S$ and $T$ share identical values, the bit-flip count in the X-Z coordinate exceeds that in the Y-Z coordinate. This observation leads to the conclusion that the greater the distance, the smaller the impact on the victim row.

\vspace{0.05em}
\fbox{\begin{minipage}{25em}
\textbf{Obs.\#2.} Edge aggressor rows have less influence than normal aggressor rows on the victim row. 
\end{minipage}}

For mf-A chips characterized in Fig. \ref{results1}(a), as $T$ increases, the number of bit-flips remains nearly constant, indicating that the Threshold Breaker is unable to decrease $S$ by elevating $T$. The distinction lies in Fig. \ref{results1}(b), where with the rise in $T$, the increment in the number of bit-flips surpasses that of increasing $S$. However, a notable issue with Fig. \ref{results1}(b) is the extremely low total number of bit-flips which means mf-B is robust against any RH fault injection method including Threshold Breaker. Although the bit-flips significantly increase with the Threshold Breaker, the impact remains negligible at the chip level. 
Moreover, mf-B encounters the same challenge as both mf-E chips and mf-F chips in Fig. \ref{results1}(e)(f). They can generate a substantial number of bit-flips, and $T$ demonstrates a clear influence on bit-flips. However, both of them reach a significantly high threshold of $S$ that affects bit-flips.
In this case, bit-flips will only experience a notable increase when $T$ surpasses $S$, and this contradicts our initial hypothesis where we need neither $T$ nor $S$ to be higher than the HC needed for the double-sided model otherwise, the Threshold Breaker will get detected. Therefore, at the high level, we can divide the under-test chips into two categories, i.e., Failed Bypass and Bypass as shown in Fig. \ref{results1}.

\fbox{\begin{minipage}{25em}
\textbf{Obs.\#3.}  Our evaluation exhibits that 50\% of the modules have much stronger resistance to RH-based bit-flip attacks including our proposed Threshold Breaker.
\end{minipage}}

The points where the planes intersect with the plot represent all HCs capable of producing that specific quantity of bit-flips. Taking mf-H characterized in Fig. \ref{results1}(h) as an example, the double-sided model's HC set is (2M,0) and the ARVRA model's HC set is (0,8M). Excluding the two edge sets, we can identify the optimal HC set along this intersection line. This HC set must satisfy the following criteria: $(i)$ Both $S$ and $T$ must be smaller than $T$ in the double-sided model, which is 2 million in the aforementioned (2M, 0). $(ii)$ The values of $S$ and $T$ should be as close or even equal as possible as we call it soft-hammering. Table \ref{Samsung} provides intuitive data of mf-H to observe the comparison between configuration sets of the three models. Our observation indicates that when $S$=$T$, both $S$ and $T$ simultaneously reach the minimum HC in the AAVAA model. This implies that the Threshold Breaker can effectively bypass the counter, causing the most bit-flips with the least HC. Among Fig. \ref{results1}(c),(d),(g),(h), it's evident that mf-C yields the most favorable results. The optimal configuration set is (1M, 1M), contrasting with the double-sided model's set of (2M, 0). This signifies that the Threshold Break can adeptly evade any counter-based detection without incurring any additional overhead. In contrast, compared with the double-sided model, the total HC required for manufacturers' chips characterized in Fig. \ref{results1}(d),(g),(h) increases in the AAVAA model when bypassing counter detection. For instance, in the case of mf-D, where $S$+$T$ equals 1.8M, there's an 80\% increase compared to the double-sided model. Nonetheless, based on previous experiments, this value remains within a feasible range.
\vspace{0.2em}
\fbox{\begin{minipage}{25em}
\textbf{Obs.\#4.} The attacker can conduct a successful RH attack on certain chips with a significantly smaller HC than the double-sided model by selecting the proper set of ($S$,$T$).
\end{minipage}}
\vspace{0.2em}

We assume the threshold of the RH to approximate the maximum HC that can be inserted within $t_{REFW}$, which can be calculated with the formula in Section 3. As discussed, to acquire more comprehensive data in the experiment, we turned off the refresh window, enabling unrestricted experimentation. Experimental findings reveal that as the overall number of RH operations increases, the value of $S$+$T$ in the ideal pattern tends to approach $S$ in the double-sided model. In fact, even in the case of mf-C, the value of $S$+$T$ in the ideal pattern can be smaller than $S$ in the double-sided model.


\begin{table}[h]
\caption{Comparable sets in three models.}\vspace{-1em}
\scalebox{0.85}{
\begin{tabular}{ccc|ccc|ccc}
\multicolumn{3}{c|}{Double-sided}                                                                           & \multicolumn{3}{c|}{ARVRA}                                                         & \multicolumn{3}{c}{AAVAA}                                                               \\ \hline
S                        & T                          & Bit-flips                                           & S                         & T                        & Bit-flips                   & S                           & T                           & Bit-flips                   \\ \hline
0                        & 500k                       & \cellcolor[HTML]{FFFFFF}200                         & 500k                      & 0                        & 10                          & 500k                        & 500k                        & 215                         \\
0                        & 1M                         & \cellcolor[HTML]{FFFFFF}522                         & 1M                        & 0                        & 55                          & 900k                        & 900k                        & 514                         \\
{\color[HTML]{FE0000} 0} & {\color[HTML]{FE0000} 2M}  & {\color[HTML]{FE0000} 980}                          & 5M                        & 0                        & 752                         & {\color[HTML]{FE0000} 1.6M} & {\color[HTML]{FE0000} 1.6M} & {\color[HTML]{FE0000} 970}  \\
0                        & 5M                         & \cellcolor[HTML]{FFFFFF}1799                        & {\color[HTML]{FE0000} 8M} & {\color[HTML]{FE0000} 0} & {\color[HTML]{FE0000} 1005} & {\color[HTML]{009901} 5M}   & {\color[HTML]{009901} 5M}   & {\color[HTML]{009901} 2557} \\
{\color[HTML]{009901} 0} & {\color[HTML]{009901} 10M} & \cellcolor[HTML]{FFFFFF}{\color[HTML]{009901} 2486} & 10M                       & 0                        & 1343                        & 10M                         & 10M                         & 3850                       
\end{tabular}}
\label{Samsung}\vspace{-1em}
\end{table}

Conducting an unrestricted experiment also serves the purpose of formulating a hypothesis for DDR5 chips as a future work. It is well-studied that DDR5 requires fewer HCs to induce bit-flips, and DDR5 boasts faster read and write speeds. Hence, our framework can be used to run experiments on DDR5 by removing the refresh window of DDR4 to prolong the time during which RH attacks can occur. If our hypothesis holds, the experimental outcomes would indicate that the Threshold Breaker poses a more pronounced threat to counter-based defense mechanisms on DDR5.
\vspace{-1em}

\subsection{Case-study: Targeted Attack against DNNs}

The RH-based DNN weight attack such as Bit-Flip Attack (BFA) \cite{rakin2019bit,yao2020deephammer} progressively searches for vulnerable bits by first performing a bit ranking within each layer based on gradient. Considering a weight quantized DNN, the weight matrix can be parameterized by two's complement representations $\{\tB_l\}_{l=1}^L$, where $l \in \{1, 2,...,L\}$ is the layer index. 
BFA computes the gradient w.r.t. each bit of the model ($ |\nabla_{\tB_l} \mathcal{L}|$) where $\mathcal{L}$ is the inference loss function. At each iteration, the attacker performs two key attack steps: i) inter-layer search and ii) intra-layer search, where the goal is to identify a vulnerable weight bit and flip it. Given a sample input $x$ and label $t$, the BFA~\cite{rakin2019bit} algorithm tries to maximize the following loss function ($\mathcal{L}$):
\vspace{-8pt}
\begin{equation}
\label{eqt:BFAN}
\begin{gathered}
\max_{\{\hat{\tB}_l\}}  ~\mathcal{L}\Big (f \big( \vx ; \{\hat{\tB}_l\}_{l=1}^{L} \big), {\vt} \Big),
\end{gathered}
\vspace{-5pt}
\end{equation}

\noindent while ensuring the hamming distance between the perturbed weight tensor by BFA ($\hat{\tB}_{l=1}^L$) and initial weight tensor ($\{\tB_l\}_{l=1}^L$) remains minimum. Finally, the attack efficiency can be measured by the number of bit-flips required to cause DNN malfunction. 

\noindent \textbf{White Box Threat Models.}  At the hardware side, we assume that 1) Each row is assigned a $T_{RH}$ upon transforming into an aggressor row. If this threshold is exceeded within the $T_{ref}$, the row will induce a bit-flip in adjacent victim rows; 2) Vulnerable data rows are not concentrated in one or two sub-arrays, nor uniformly distributed across each sub-array. Experimental findings reveal that most sub-arrays concurrently store several data rows, with some potentially storing multiple or none at all; and 3) The attacker possesses a comprehensive mapping file, allowing them to pinpoint the physical address of the target data within the DNN. Additionally, they know the initial static mapping of DRAM rows, with physical adjacency information between rows \cite{jattke2022blacksmith,wi2023shadow}. Consequently, the attacker can execute an RH attack on the targeted content. At software side, we adopt a standard white-box threat model for the BFA. In this white-box threat model, the attacker possesses knowledge of the internal structure of DNN models, including details such as the number of layers and the width of each layer. Additionally, the attacker has complete information about DNN model parameters, their values, and the bit representation used for inference. This assumption is grounded in the recent progress of side-channel information leakage and the reverse-engineering of DNN models~\cite{yan2020cache}, which enables the recovery of DNN model configurations during the inference stage.

\begin{figure}[t]
\centering
\begin{subcaptionblock}{.5\linewidth}
\centering
\includegraphics[width=\linewidth]{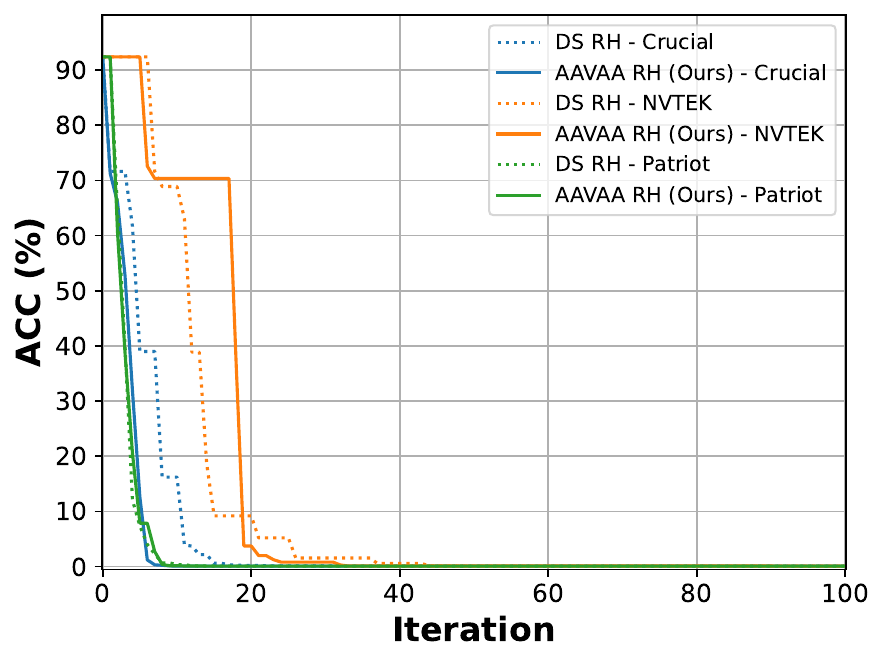}\vspace{-0.5em}
\caption{}\label{fig:imagenet-disabled}
\end{subcaptionblock}%
\begin{subcaptionblock}{.5\linewidth}
\centering
\includegraphics[width=\linewidth]{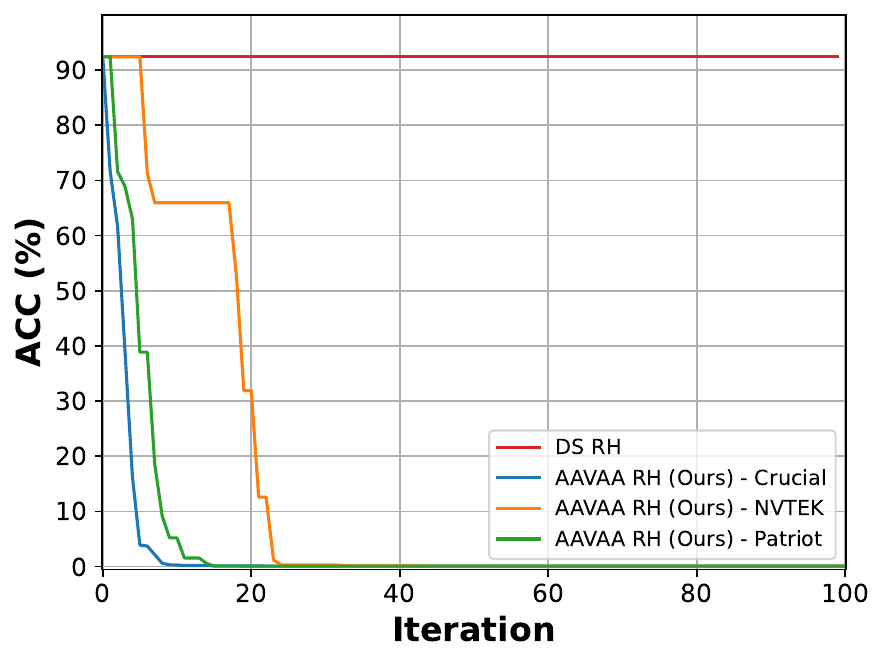}\vspace{-0.5em}
\caption{}\label{fig:imagenet-enabled}
\end{subcaptionblock}%
\vspace{-1.5em}
\caption{Comparison of double-sided (DS) and our AAVAA attack models for an 8-bit quantized ResNet-34 trained on ImageNet, with counter-based RH protection mechanisms (a) disabled and (b) enabled, respectively.} \vspace{-1em}
\label{fig:imagenet}
\end{figure}


\noindent \textbf{Results.} We compare our Threshold Breaker's AAVAA fault injection model with the traditional double-sided attack on three different DRAM chips for an 8-bit quantized ResNet-34~\cite{he2015delving} trained on ImageNet~\cite{deng2009imagenet} (as shown in Fig.~\ref{fig:imagenet}). We follow the weight quantization scheme in~\cite{rakin2019bit}. 
While the AAVAA model doesn't exhibit a clear improvement in attack efficiency with disabled counter-based RH protection mechanisms, it successfully circumvents these protections when enabled, unlike the double-sided RH attack. This underscores the Threshold Breaker's ability to effectively bypass counters, posing a significant threat to the safety of deep learning applications, especially those critical in various scales. The findings highlight the vulnerability of counter-based RH defenses, emphasizing the need for robust strategies to mitigate the risks posed by novel fault injection techniques in safety-critical systems reliant on deep learning technologies.

\section{Conclusion}
The paper introduces a new fault injection attack, the Threshold Breaker, challenging RowHammer detection methods. By soft-attacking rows at a greater physical distance from the target, this technique effectively bypasses advanced counter-based defenses. While no prior work has demonstrated the effect of such an attack, our work closes this gap by systematically testing 128 real commercial DDR4 DRAM products, revealing its impact on chips from major manufacturers. A case study further demonstrated the superiority of the adversarial weight attack leveraging Threshold Breaker on a DNN trained on ImageNet while DRAM remains fully protected compared to the well-known double-sided RowHammer model. By diminishing the size of DRAM chips through modern manufacturing processes, DRAM becomes progressively more susceptible to RowHammer bit-flips. We believe that the Threshold Breaker would represent a more pronounced threat to counter-based defense mechanisms on DDR5.

\section{Acknowledgment}
We would like to extend our gratitude to Professor Onur Mutlu and Ataberk Olgun at ETH Zurich for generously sharing their expertise and providing valuable insights into the DRAMBender infrastructure \cite{olgun2023dram}  to conduct this study.

\bibliographystyle{ACM-Reference-Format}
\bibliography{Main}\vspace{-2em}

\end{document}